\documentclass[aps,prl,reprint,groupedaddress]{revtex4-2}

\usepackage[margin=1.2in]{geometry}
\usepackage[T1]{fontenc}
\usepackage[utf8]{inputenc}
\usepackage{lmodern}
\usepackage{hyperref}
\usepackage{graphicx}
\usepackage[english]{babel}

\usepackage{siunitx}

\usepackage{color,soul}
\usepackage{bigints}

\usepackage{newunicodechar}
\newunicodechar{ﬁ}{fi}
\newunicodechar{ﬀ}{ff}

\usepackage{hyperref}
\usepackage{graphicx}
\usepackage[english]{babel}
\usepackage{color,soul}
\usepackage{bigints}
\usepackage{graphicx}
\usepackage{float}
\usepackage{tabu}
\usepackage{tabularx}
\usepackage{booktabs}
\usepackage{lipsum}
\usepackage{mwe}
\usepackage{url}
\usepackage{hyperref}
\hypersetup{pdftex,colorlinks=true,allcolors=blue}
\usepackage{hypcap}
\usepackage{amssymb}
\usepackage{caption}
\usepackage{subcaption}
\usepackage{xfrac}
\usepackage{comment}

\usepackage{diagbox}

\usepackage{blindtext}
\usepackage{latexsym}
\usepackage{wasysym}\usepackage{ amssymb }
\usepackage{xcolor}
\usepackage{color}
\usepackage{lipsum}

\definecolor{blue-matlab}{rgb}{0,0.4470,0.7410}
\definecolor{orange-matlab}{rgb}{0.8500,0.3250,0.0980}
\definecolor{yellow-matlab}{rgb}{0.9290,0.6940,0.1250}
\definecolor{purple-matlab}{rgb}{0.4940,0.1840,0.5560}

\definecolor{blue-rawad}{rgb}{0.2,0.9,1}

\usepackage{color, soul}

\usepackage{lipsum}

\usepackage[framemethod=tikz]{mdframed}

\usepackage{bigints}
\usepackage{relsize}
\usepackage[font={small}]{caption}

\usepackage{tikz}
\tikzset{mynode/.style={draw,solid,circle,inner sep=1pt}}

\usepackage[font={small}]{caption}

\begin{document}


\title{Chaos in a melting pot}


\author{Rawad Himo~}\email[Rawad Himo]{Rawad.Himo@univ-nantes.fr}

\author{Cathy Castelain~}\email[Cathy Castelain]{Cathy.Castelain@univ-nantes.fr}
\author{Teodor Burghelea~}\email[Teodor Burghelea]{Teodor.Burghelea@univ-nantes.fr}
\affiliation{Universit\'{e} de Nantes, CNRS, Laboratoire de thermique et \'{e}nergie de Nantes, LTeN, UMR 6607, F-44000 Nantes, France.}


\date{\today}

\begin{abstract}

A novel flow instability emerging during a rheometric flow of a phase change material sheared in the vicinity of the solid-fluid transition is reported. Right above the onset of the flow induced crystallisation, the presence of the crystals in the flow leads to a primary bifurcation towards an oscillatory flow state. A  further decrease of the temperature beyond this point leads to an increase of the both the volume fraction and the size of the crystals which ultimately triggers a fully developed chaotic flow. A full stability diagram as a function of the imposed rate of deformation and the temperature is obtained experimentally. The experimental findings are complemented by a simple numerical toy model which, consistently with the experimental observations, indicates that the primary bifurcation is a second order bifurcation that can be accurately described by the stationary Landau-Ginzburg equation with a field.

\end{abstract}


\maketitle

In the absence of inertial contributions, a hydrodynamic system is still prone to losing its hydrodynamic stability when a physical quantity contributing to the momentum balance becomes strongly stratified in space. To help illustrate this point, thermal convection may be triggered by differentially heating a flow cavity from below \cite{RBoriginal} or gravity induced density stratification may sustain internal gravity waves, \cite{Landau1987}. The loss of hydrodynamic stability due to a viscosity stratification has been predicted theoretically several decades ago \cite{yih_1967,Hickox1971,Govindarajan2014,Valluri2010}. More recently, it has been demonstrated experimentally that linear, laminar and steady shear flows lose their stability in the presence of a strong stratification of the viscosity, \cite{doi:10.1063/1.2759190}. Phase changing materials represent a broad class of materials that undergo a solid-liquid phase transition when the temperature is gradually decreased. Although the flows of such materials are ubiquitous in many polymer processing operations including (but not limited to) film casting, melt blowing, thermoforming, their hydrodynamic stability has received practically no attention. 

\begin{figure}
\centering
\includegraphics[height=0.22\textheight]{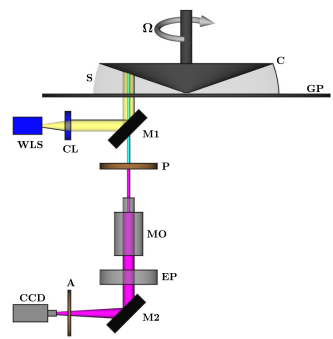} 
\captionsetup{justification=centering}
\caption{Schematic representation of the rheometer setup (not in scale): \textbf{(C)} - cone, \textbf{(GP)} - glass plate, \textbf{(S)} - sample, \textbf{(WLS)} - white light source, \textbf{(CL)} - collimating lens, \textbf{$\mathbf{(M_1)}$} - semitransparent mirror, \textbf{$\mathbf{(M_2)}$} - plane mirror, \textbf{(P)} - polariser, \textbf{(MO)} - microscope objective, \textbf{(CCD)} - charged-coupled device, \textbf{(EP)} - eye piece, \textbf{(A)} - analizer.}
\label{figRheometer}
\end{figure}

We report in this letter a novel instability observed in a low Reynolds number rheometric flow of a paraffin wax when the temperature is gradually decreased past the onset of the crystallisation. The experimental setup is schematically illustrated in Fig.\ref{figRheometer}. It consists of a $60~mm$ diameter and $2~deg$ angle cone mounted on a commercial rheometer (Mars III, ThermoFischer Scientific). Simultaneously with the classical macro-rheological tests of the apparent viscosity $\eta$, the micro-structure of the material is visualised through crossed polarisers using a microscope mounted bellow the bottom plate of the setup,  Fig.\ref{figRheometer}.

\begin{figure*}
\centering
\includegraphics[height=0.22\textheight]{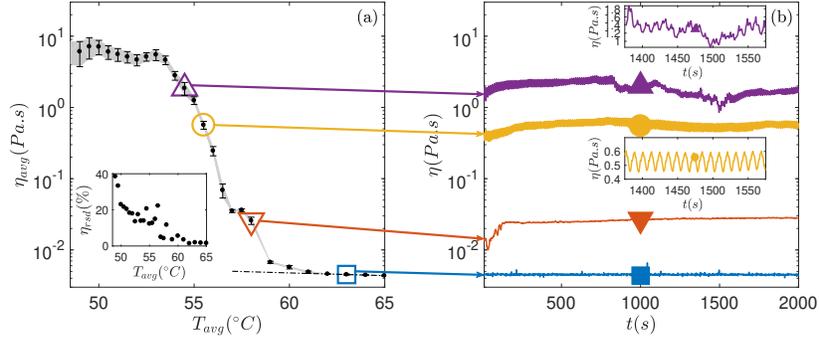} 
\captionsetup{justification=centering}
\caption{(a) Average apparent viscosity measured isothermally at a constant shear rate $\dot{\gamma}=15~(1/s)$. Full markers • represent averages over $2000~s$ and the error bars $\mathrm{I}$ are defined by the standard deviation. The temperature dependence of the standard deviation is presented in the insert. (b): Viscosity time series showing four distinct phases. The empty symbols in panel (a) refer to the flow states:  $\begingroup \color{blue-matlab} \blacksquare \endgroup$ $(T=63^{\circ} C)$ Laminar,   $\begingroup \color{orange-matlab}  \blacktriangledown \endgroup$ $(T=58^{\circ} C)$ Crystal formation,  $\begingroup \textcolor{yellow-matlab} \CIRCLE \endgroup$ $(T=55.5^{\circ} C)$ Periodic Behavior,  $\begingroup \textcolor{purple-matlab} \blacktriangle  \endgroup$ $(T=54.5^{\circ} C)$ Chaos. }
\label{figEtaTimeSeries}
\end{figure*}

Subsequent to reaching temperature equilibrium with a precision of $0.01^oC$ during $200~s$, measurements of the apparent shear viscosity averaged during $2000~s$ performed at a constant rate of shear $\dot{\gamma}=15~s^{-1}$ and various temperatures are presented in Fig. \ref{figEtaTimeSeries}(a).  In a fluid regime ($T>61^oC$) the time averaged viscosity follows an Arrhenius dependence, $\eta= \left( 1.309 \pm 0.758 \right) \times 10^{-5} exp \left( \frac{1964 \pm 195}{T  } \right) $. In this regime the instrumental error of the viscosity measurements does not exceed $2\%$ of the mean value.  Upon a gradual decrease of the temperature past the fluid regime a sharp increase of  the apparent viscosity is observed. This corresponds to the onset of crystallisation. Upon a further decrease of the temperature the time averaged apparent viscosity increases up to two orders of magnitude.   A rather intriguing feature observed within this range of temperatures relates to the level of fluctuations of the apparent viscosity which has increased drastically up to $5\%$ of the mean value, panel (a) in Fig.\ref{figEtaTimeSeries}(a). As according to the calibration data of the rheometer within this range of torques the instrumental error does not exceed $2\%$ of the mean value, the possibility of spurious torque measurements can be safely ruled out.

To get further insights into the dynamics of the liquid-solid transition, we focus on individual time series of the apparent viscosity, Fig. \ref{figEtaTimeSeries}(b). Whereas at $T=63^{\circ} C$ the apparent viscosity time series shows no fluctuations other than those related to the instrumental error - the series marked by a solid square,  a monotonic increase is observed at $T=58^{\circ} C$ - the time series labeled by a solid down triangle. This may be attributed to the onset of crystallisation. At this temperature, however, the apparent viscosity time series exhibits no fluctuations which is consistent with a hydrodynamically stable flow regime and indicates that the crystals did not grow in size sufficiently in order to destabilise the flow. 
At a slightly lower temperature, $T=55.5^{\circ} C$, the time series of the apparent viscosity becomes oscillatory indicating a loss of the hydrodynamic stability - the time series marked by a full circle. 
Within this regime, the paraffin crystals became sufficiently large in order to destabilise the flow and the solid and fluid elements coexist, Fig. \ref{figEtaTimeSeries}(b).
Finally, at $T=54.5^{\circ} C$ the apparent viscosity evolves chaotically in time - the time series labeled by a solid up triangle.

\begin{figure}
\centering
\includegraphics[height=0.2\textheight]{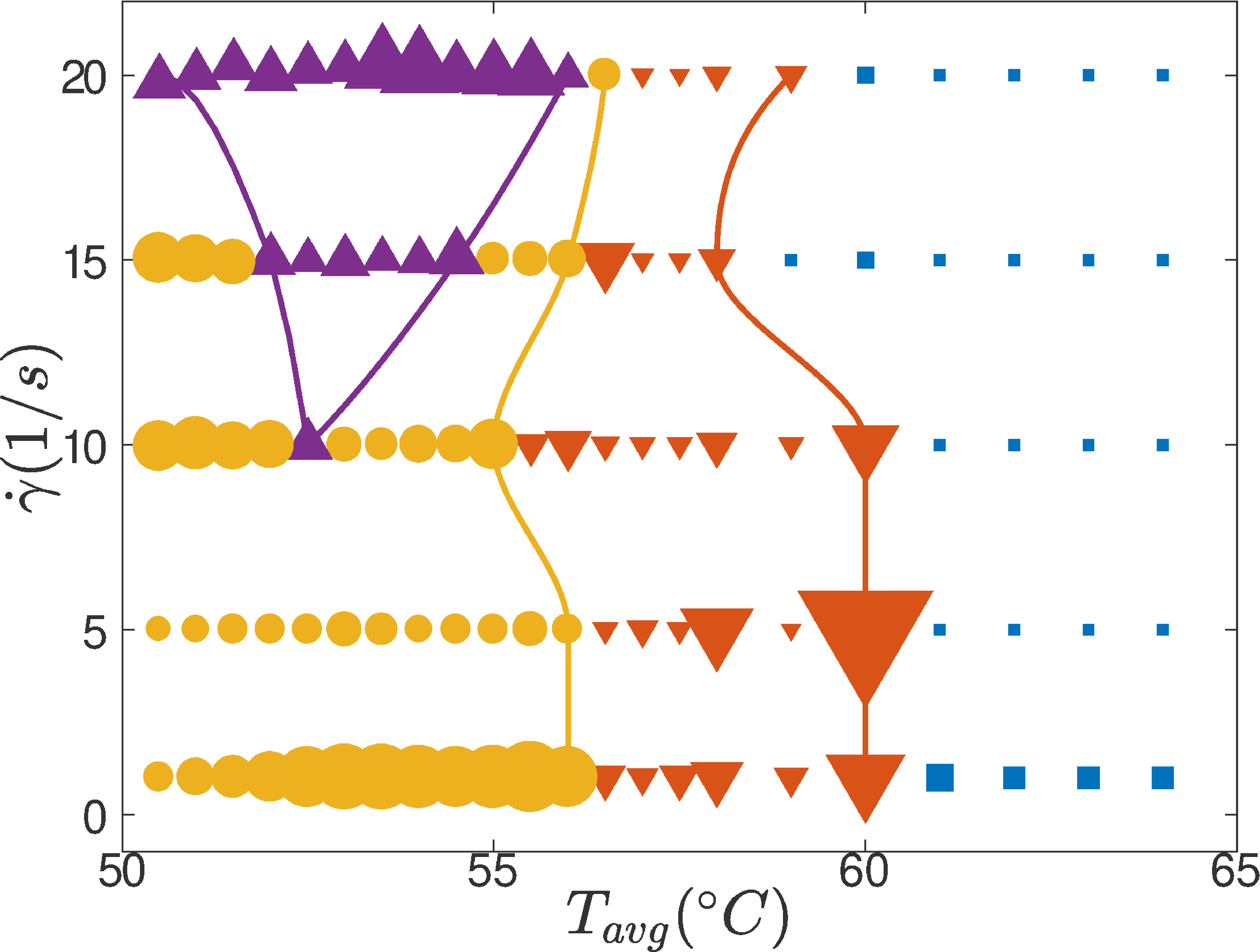} 
\captionsetup{justification=centering}
\caption{Hydrodynamic stability diagram:  $\begingroup \color{blue-matlab} \blacksquare \endgroup$ - stable,  $\begingroup \color{orange-matlab}  \blacktriangledown \endgroup$ - stable, crystal formation,  $\begingroup \textcolor{yellow-matlab} \CIRCLE \endgroup$ - oscillatory,  $\begingroup \textcolor{purple-matlab} \blacktriangle  \endgroup$ - chaotic. The size of the symbols is proportional to the relative standard deviation (rsd) of fluctuations of the apparent viscosity.}
\label{figParaffinChoasPhaseDiagram}
\end{figure}

To obtain a full picture of the hydrodynamic stability of the system measurements similar to those illustrated in Fig. \ref{figEtaTimeSeries} have been performed for several values of the imposed shear rate. The results are summarised in the stability diagram presented in Fig. \ref{figParaffinChoasPhaseDiagram}.

Regardless the value of the shear rate, a monotonic increase of the apparent viscosity related to onset of the crystallisation in the flow (the triangles) is observed within this regime which, upon a further decrease of the temperature, is followed by an oscillatory instability (the circles). 

\begin{figure}
\centering
\begin{tabular}{cc}
\includegraphics[trim=70 90 70 90,clip,height=0.135\textwidth , width=0.18\textwidth]{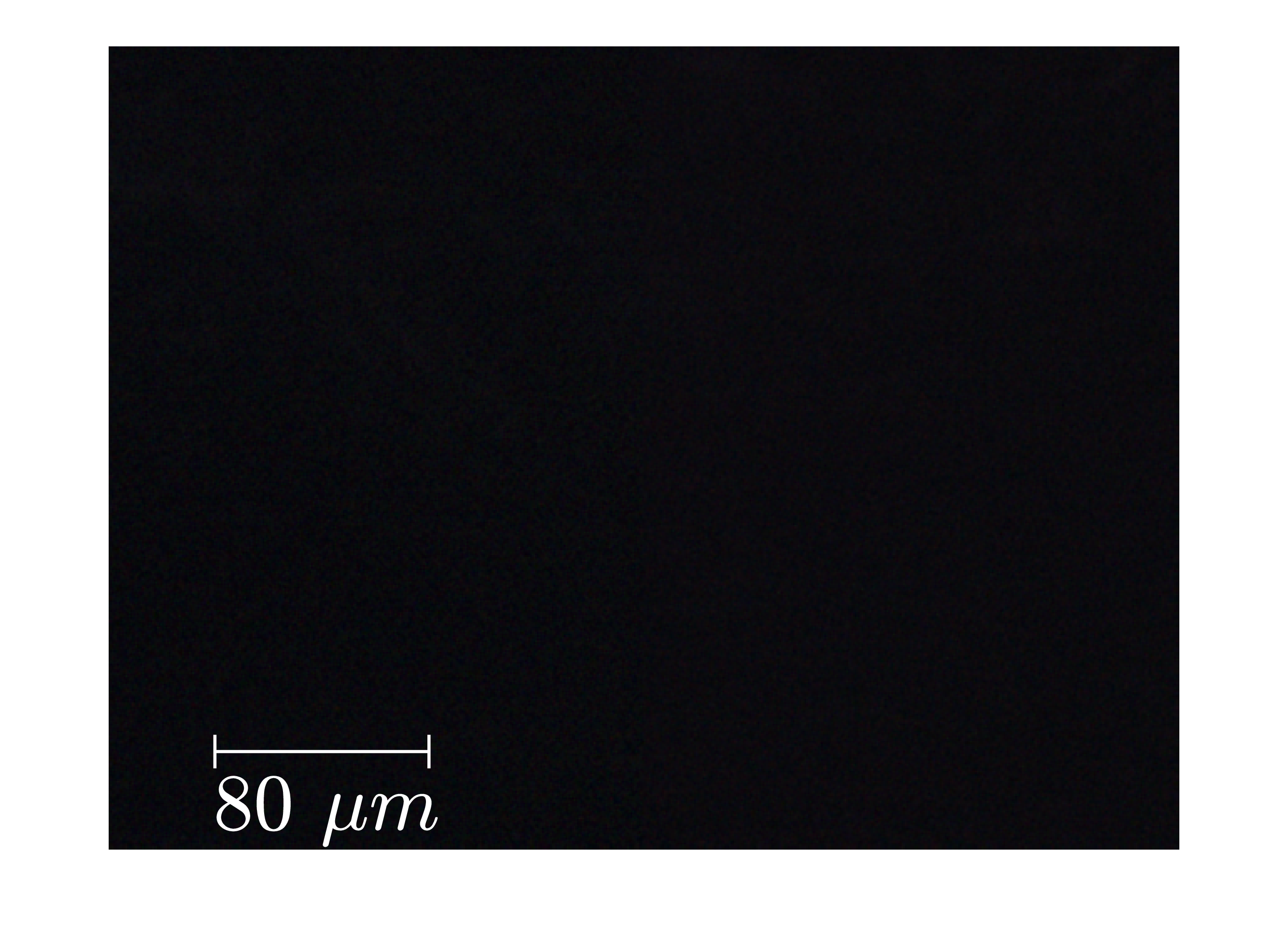} &  \includegraphics[trim=70 90 70 90,clip,height=0.135\textwidth , width=0.18\textwidth]{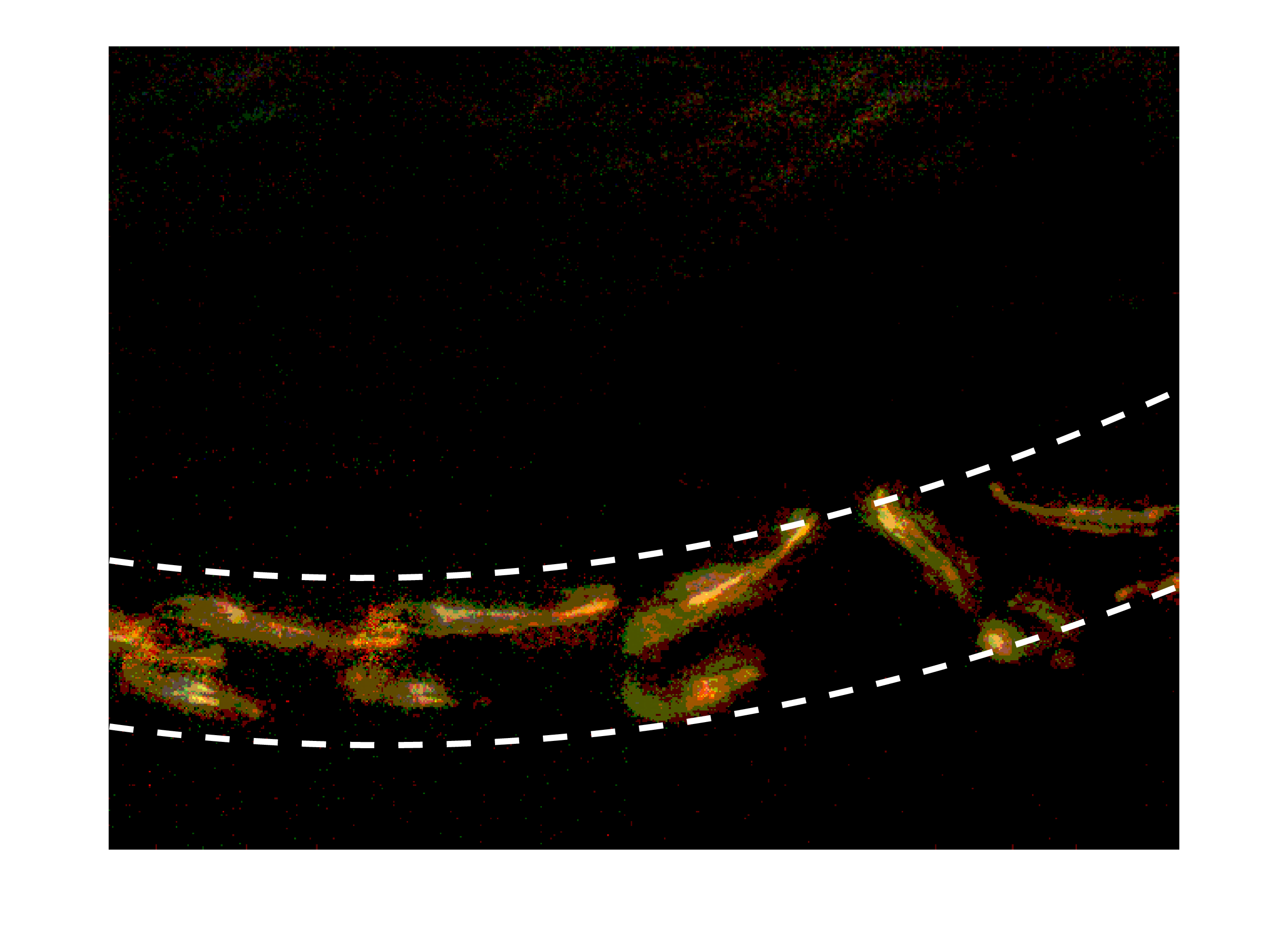} \\
\small \textbf{(a)} $T=63^{\circ}C$ & \small \textbf{(b)} $T=58^{\circ}C$ \\
 \includegraphics[trim=70 90 70 90,clip,height=0.135\textwidth , width=0.18\textwidth]{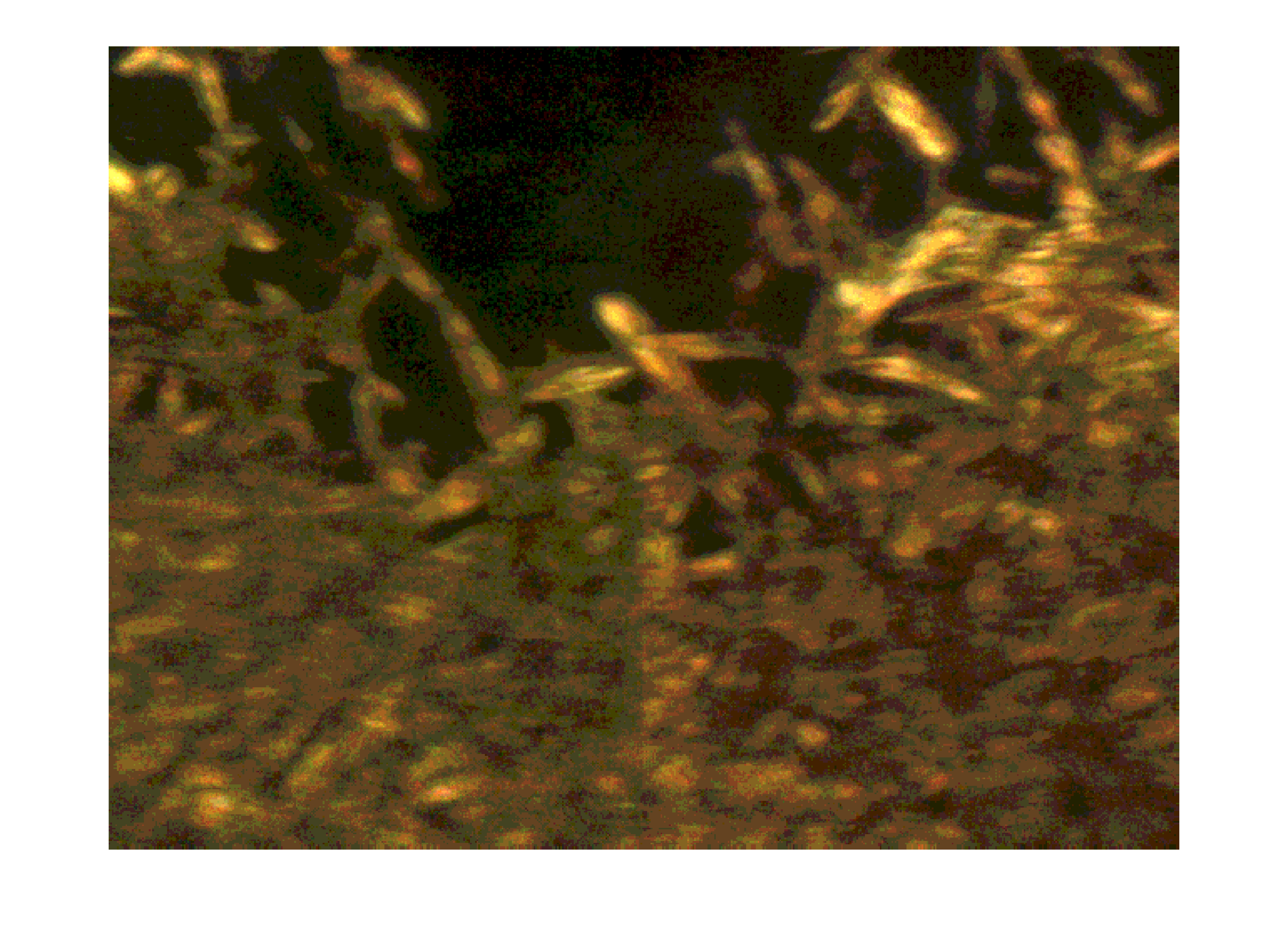} &  \includegraphics[trim=70 90 70 90,clip,height=0.135\textwidth , width=0.18\textwidth]{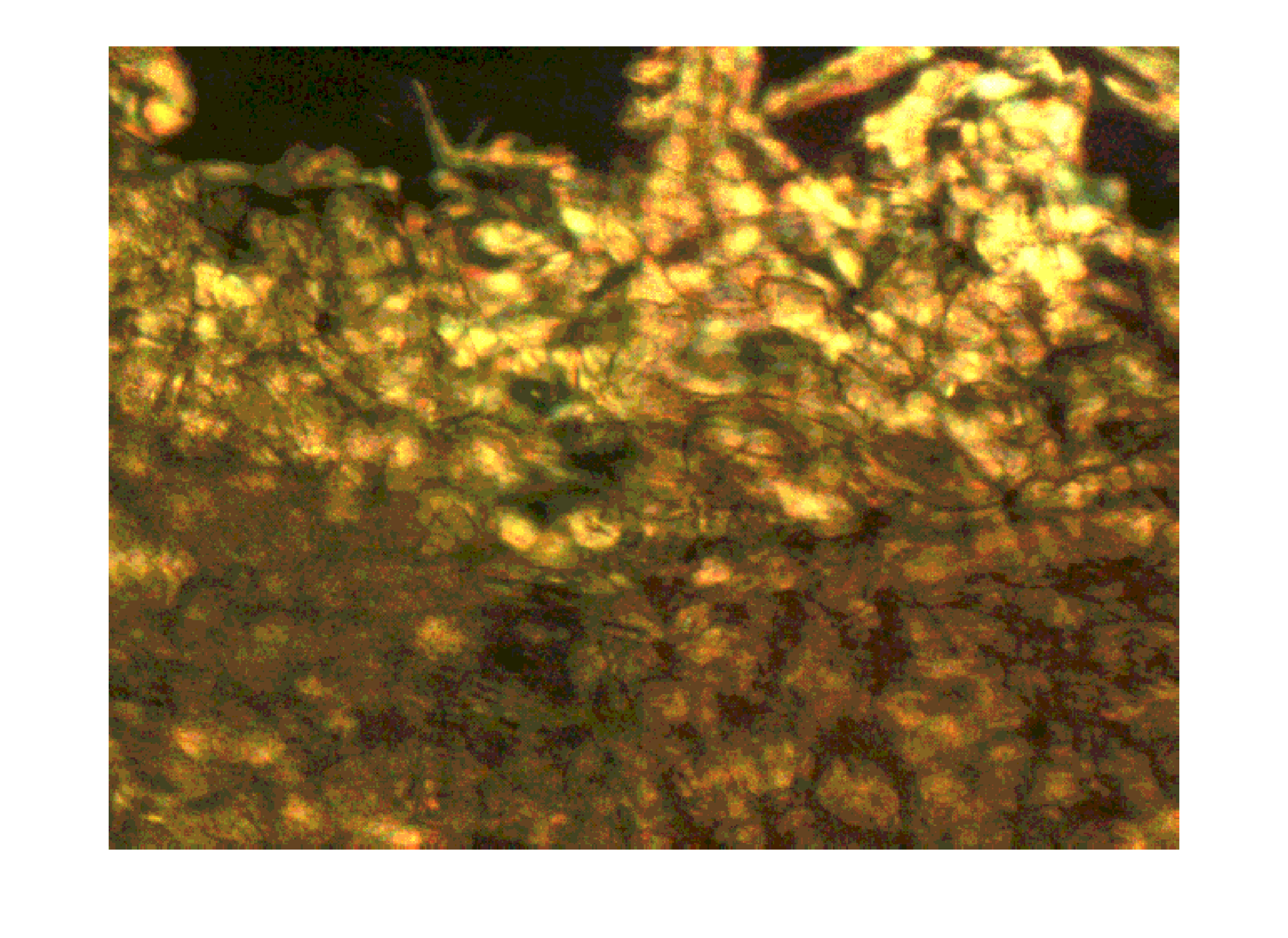} \\
\small \textbf{(c)} $T=55.5^{\circ}C$ & \small \textbf{(d)} $T=54.5^{\circ}C$\\
\end{tabular} 
\captionsetup{justification=centering}
\caption{Polarised light micrographs of the flow acquired at a constant shear rate a $\dot \gamma =15 s^{-1}$ and  several temperatures indicated in the lower inserts. The size of the field of view is $400 \mu m$.}
\label{figRheometerMicroscopy}
\end{figure}

To understand the physical origins of both the primary oscillatory instability and the ultimate chaotic behaviour observed during the macro-rheological  
measurements presented in Fig. \ref{figEtaTimeSeries} we resort to an in-situ visualisation of the micro structure using a polarised microscope mounted underneath the bottom plate of the rheometer as schematically illustrated in Fig. \ref{figRheometer}. Micro-graphs of the sample acquired at several temperatures and a constant shear rate $\dot \gamma =15~s^{-1}$ are presented in Fig. \ref{figRheometerMicroscopy}.    
At $T=58^{\circ}C$ which corresponds to the monotonic and steady increase of the apparent viscosity  (the triangles in Fig. \ref{figParaffinChoasPhaseDiagram}) few small paraffin crystals following a roughly circular path may be observed in the field of view (panel (b) in Fig. \ref{figRheometerMicroscopy}). At $T=55.5^{\circ}C$ which corresponds to the oscillatory flow regime marked by full circles in Fig. \ref{figParaffinChoasPhaseDiagram} a coexistence between large aggregates of crystals and fluid paraffin is observed.  Decreasing the temperature even further to $T=54.5^{\circ}C$ leads to a chaotic motion of the solid-fluid mixture. The in-situ visualisation of the two phase flow indicates that the primary oscillatory instability and the later bifurcation towards a chaotic flow regime are related to the presence of the crystals. When these crystals become large enough, their presence leads to a strong stratification of the viscosity which ultimately destabilises the flow.

\begin{figure}
\centering
\begin{tabular}{ccc}
\multicolumn{3}{c}{\includegraphics[height=0.04\textwidth]{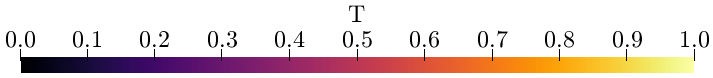}} \\
\includegraphics[height=0.12\textwidth]{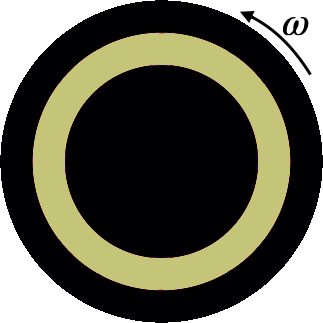} ~~&  \includegraphics[height=0.12\textwidth]{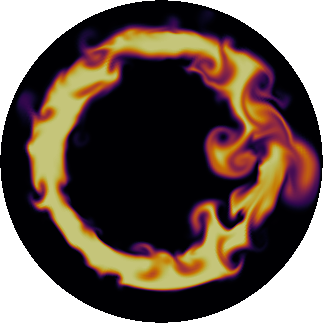} ~~&  \includegraphics[height=0.12\textwidth]{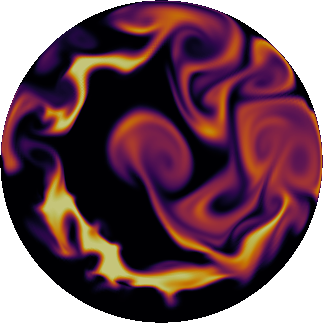} \\
\scriptsize \textbf{(a)} $t^*=0$ &\scriptsize \textbf{(b)} $t^*=0.0046$ & \scriptsize \textbf{(c)} $t^*=0.0094$ \\
\end{tabular} 
\captionsetup{justification=centering}
\caption{Scalar transport contours for $Re=1$ and $\chi=100$ at different time steps.}
\label{figOpenFOAMRe1Nu100}
\end{figure}

To test this phenomenological picture, a simple toy-model is proposed. A viscous fluid confined within a disk is set in motion by rotating the circumference of the disk at a constant angular speed $\Omega$. The viscosity stratification is modelled by considering that within a thin circular ring, the viscosity is $\chi$ times larger than the viscosity outside the ring, Fig. \ref{figOpenFOAMRe1Nu100}(a). This configuration of the base flow mimics well the onset of crystallisation visualised via polarised microscopy, Fig. \ref{figRheometerMicroscopy}(b). The angular speed of rotation of the boundary was chosen such as the Reynolds number does not exceed $Re=0.0575$ which is the largest value attained during the experiments. The numerical simulations have been implemented under the open source code OpenFOAM and, for the technical details, the reader is referred to the Supplemental Material. 

\begin{figure}
\centering
\includegraphics[height=0.2\textheight]{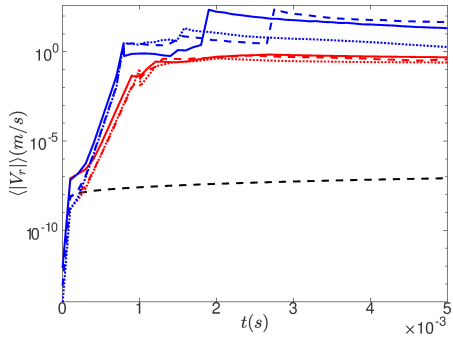} 
\captionsetup{justification=centering}
\caption{Area weighted average of the absolute radial velocity $\langle |V_r| \rangle_r$  versus time, for several Reynolds numbers: dotted ($\cdot  \cdot  \cdot$) for $Re=0.1$, dashed (- -) for $Re=1$, and solid lines (\textemdash) for $Re=10$. The colours refer to the viscosity ratio: black for $\chi=10$, \textcolor{red}{red for $\chi=100$} and \textcolor{blue}{blue for $\chi=1000$}.}
\label{figVrVsTimeReynolds}
\end{figure}

Corresponding to a viscosity ratio $\chi=100$ the base flow loses its hydrodynamic stability in a finite time, $t^*=0.0046$ - panel (b) in Fig. \ref{figOpenFOAMRe1Nu100} and gets increasingly chaotic as the time elapses panel (c) in Fig. \ref{figOpenFOAMRe1Nu100}. In order to verify the hypothesis that this instability is triggered by the viscosity contrast rather than the inertia,  time series of the space averaged radial velocity component   $\langle |V_r| \rangle$ computed for three values of viscosity ratio ($\chi=10,~100,~1000$) and three values of the Reynolds number ($Re=0.1,~1,~10$) are presented in Figure~\ref{figVrVsTimeReynolds}. Whereas the viscosity ratio clearly influences the evolution of the space averaged radial velocity, the Reynolds number plays no role which is a direct confirmation of the fact that the instability is triggered by the spatially inhomogeneous distribution of the viscosity in the base flow.

\begin{figure}
\centering
\includegraphics[height=0.3\textwidth]{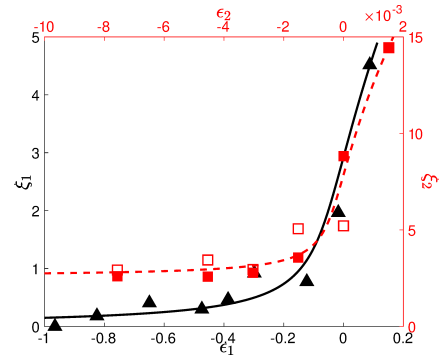} 
\captionsetup{justification=centering}
\caption{Left-bottom, up-triangles: Dependence of the order parameter $\chi_1$ on the control parameter $\epsilon_1$ obtained from the numerical simulations (see text for description). The full line is a fit by the stationary Landau-Ginzburg equation. Right -top, squares: Dependence of the order parameter $\xi_2$ on the control parameter $\epsilon_2$ obtained from macro-rheological experiments (see text for description). The full/empty symbols correspond to the increasing/decreasing branch of the temperature ramp. The data were acquired at a fixed imposed shear rate $\dot \gamma=5~s^{-1}$. The dashed line is a fit by the stationary Landau-Ginzburg equation.}
\label{figLandau}
\end{figure}

We now focus on understanding the physical nature of the primary bifurcation towards the oscillatory flow states labelled by full circles in Fig.  \ref{figParaffinChoasPhaseDiagram}. First, the reduced standard deviation (std) of the level of fluctuations of the apparent viscosity $\xi_2 = \frac{\eta^{std}}{\eta_{avg}}$ is measured during a decreasing/increasing ramp of temperatures is monitored versus the reduced temperature $\epsilon_2= \frac{T}{T_m}-1$ - the full/empty squares in Fig. \ref{figLandau} where $T_m \approx 57.25 ^oC$ is the melting temperature measured via Differential Scanning Calorimetry (DSC). The reduced level of fluctuations is reversible upon decreasing/increasing reduced temperature and can be described by the stationary Landau-Ginzburg equation with a field - the full line. To compare with the instability observed via numerical simulations, the dependence of the terminal value of the space averaged magnitude of the secondary flow $\xi_1=\left< \langle |V_r| \rangle_r\right>_t$ on the reduced viscosity ratio $\epsilon_1=\frac{\chi}{\chi_c}-1$ is shown in the same plot - the top-bottom axes and the triangles.  The dependence of the order parameter $\xi_1$ on the reduced controlled parameter $\epsilon_1$ obtained from the numerical simulations is well fitted by the stationary  Landau-Ginzburg
equation.

Based on both the experimental measurements and the results of the numerical simulations presented in Fig. \ref{figLandau}, one may conclude that the transition towards the oscillatory flow states emerges as an imperfect bifurcation. The abrupt increase of the level of fluctuations of the apparent viscosity associated to the transition to a chaotic flow regime (the insert in Fig. \ref{figEtaTimeSeries}(a)) is an indicator that this second bifurcation might be first order but this remains to be clarified by future studies.  

In closing, the relevance of the present study is two-fold. From a fundamental perspective our findings clearly call for future theoretical developments on flows of phase-change materials. From a practical standpoint, such instabilities may exist in a variety of basic polymer processing operations and ultimately influence the quality of the final products.

\bibliographystyle{apsrev4-2}
\bibliography{Refs.bib}   

\end{document}